\documentclass[usenatbib]{mn2e}

\newcommand{\kms}{km~s$^{-1}$}

\newcommand{\etal}{{et al.}}
\newcommand{\ie}{{\it i.e.}}

\newcommand{\be}{\begin{equation}}
\newcommand{\ee}{\end{equation}}

\newcommand{\kmsMpc}{km~s$^{-1}$ Mpc$^{-1}$}

\newcommand{\mnras}{MNRAS}
\newcommand{\apj}{ApJ}
\newcommand{\apjl}{ApJL}
\newcommand{\aj}{AJ}
\newcommand{\pasp}{PASP}

\newcommand{\aap}{A\&A}
\newcommand{\apjs}{ApJS}
\newcommand{\araa}{ARA\&A}
\newcommand{\pbar}{p_{\rm bar}}
\newcommand{\fHI}{f_{\rm HI}}

\usepackage{epsfig}
\begin{document}

\title[Atomic gas and bars in disc galaxies]{Galaxy Zoo and ALFALFA: Atomic Gas and the Regulation of Star Formation in Barred Disc Galaxies}
\author[K.L. Masters \etal]{Karen L. Masters$^{1,2}$, Robert C. Nichol$^{1,2}$, Martha P. Haynes$^3$, William C. Keel$^4$, \newauthor Chris Lintott$^5$, Brooke Simmons$^{5,6,7}$, Ramin Skibba$^8$, Steven Bamford$^9$, \newauthor Riccardo Giovanelli$^3$ and Kevin Schawinski$^{6,7,10}$ \\
$^1$Institute for Cosmology and Gravitation, University of Portsmouth, Dennis Sciama Building, Burnaby Road, Portsmouth, PO1 3FX, UK \\
$^2$South East Physics Network, www.sepnet.ac.uk\\
$^3$Center for Radiophysics and Space Research, Cornell University, Space Sciences Building, Ithaca, NY 14850, USA\\
$^4$Department of Physics \& Astronomy, 206 Gallalee Hall, 514 University Blvd., University of Alabama, Tuscaloosa, AL 35487-0234, USA\\
$^{5}$Oxford Astrophysics, Department of Physics, University of Oxford, Denys Wilkinson Building, Keble Road, Oxford, OX1 3RH, UK\\
$^{6}$Yale Center for Astronomy and Astrophysics, Yale University, P.O. Box 208121, New Haven, CT 06520, USA \\
$^{7}$Department of Astronomy, Yale University, New Haven, CT 06511 USA \\
$^8$Steward Observatory, University of Arizona, 933 N. Cherry Ave, Tuscon, AZ 85721, USA\\
$^9$School of Physics \& Astronomy, University of Nottingham, University Park, Nottingham, NG7 2RD, UK\\
$^{10}$Einstein Fellow\\ 
\\
 $^*$This publication has been made possible by the participation of more than 200,000 volunteers in the Galaxy Zoo project. \\ Their contributions are individually acknowledged at \texttt{www.galaxyzoo.org/volunteers}. \\
\\
{\tt E-mail: karen.masters@port.ac.uk}
 }

\date{Accepted by MNRAS, 23rd May 2012}
\pagerange{1--16} \pubyear{2012}

\maketitle

\begin{abstract}
We study the observed correlation between atomic gas content and the likelihood of hosting a large scale bar in a sample of 2090 disc galaxies. Such a test has never been done before on this scale. We use data on morphologies from the Galaxy Zoo project and information on the galaxies' HI content from the ALFALFA blind HI survey. Our main result is that the bar fraction is significantly lower among gas rich disc galaxies than gas poor ones. This is not explained by known trends for more massive (stellar) and redder disc galaxies to host more bars and have lower gas fractions: we still see at fixed stellar mass a residual correlation between gas content and bar fraction.
We discuss three possible causal explanations: (1) bars in disc galaxies cause atomic gas to be used up more quickly, (2) increasing the atomic gas content in a disc galaxy inhibits bar formation, and (3) bar fraction and gas content are both driven by correlation with environmental effects (e.g. tidal triggering of bars, combined with strangulation removing gas). All three explanations are consistent with the observed correlations.  In addition our observations suggest bars may reduce or halt star formation in the outer parts of discs by holding back the infall of external gas beyond bar co-rotation, reddening the global colours of barred disc galaxies. This suggests that secular evolution driven by the exchange of angular momentum between stars in the bar, and gas in the disc, acts as a feedback mechanism to regulate star formation in intermediate mass disc galaxies. 
\end{abstract}


\section{Introduction}

 There is a growing body of evidence which suggests that secular evolution plays a vital role in the evolution of the galaxy population. Secular evolution refers to any slow processes that changes the properties of galaxies, and is often, but not exclusively, driven by internal dynamics (e.g. as the term was first used by Kormendy 1979). Observational evidence demonstrating the need for significant amounts of secular evolution is growing (e.g. Oesch et al. 2010; Cisternas et al. 2011; Schawinski et al. 2011) and theoretical models of galaxy formation are now considering its impact (e.g. Boissier \& Pratzos 2000, Debattista et al. 2006, Agertz, Teyssier \& Moore 2011, de Lucia et al. 2011, Sales et al. 2012).  Several studies now suggest that major mergers are not frequent enough, nor create the appropriate morphological transformations to be the dominant process driving galaxy evolution (e.g. Robaina et al. 2010, Bournaud et al. 2011). As the alternative mechanism, secular evolution such as minor mergers and/or gradual gas inflow must then be more important. 
 
  The strongest drivers of internal secular evolution in disc galaxies are the ``disc instabilities" known as stellar bars, (for a recent comprehensive review of both theoretical and observation status of bar studies see Section 9 of Sellwood 2010; also see Sellwood \& Wilkinson 1993). A bar, particularly a ``strong" bar, breaks the radial symmetry of the disc, allowing for the transfer of angular momentum between components (stars, dark matter and gas) and potentially driving material both inwards and outwards in the disk. As such, bars have long been invoked as a way to fuel central star formation by driving gas towards the inner regions of galaxies where it is available to fuel active galactic nuclei (AGN; probably via inner secondary bars or spiral arms e.g. Ann \& Thaker 2005), and to grow central (pseudo)-bulges (e.g. Kormendy \& Kennicutt 2004, Heller, Shloshman \& Athanassoula 2007). Observational evidence for an increase in central star formation in barred galaxies seems clear (e.g. Ho et al. 1997, Sheth et al. 2005, Ellison et al. 2011, Coehlo \& Gadotti et al. 2011, Oh, Oh \& Yi 2012, Lee et al. 2012), although the link between galactic scale bars and AGN is more controversial (e.g. Ho et al. 1997, or Oh et al. 2012, Cardamone et al. in prep.). 

Theoretical considerations suggest that the gaseous component should play a major role in the dynamics of, and the exchange of angular momentum (AM) in disk galaxies (Athanassoula 2003; Combes 2008). When a galaxy is rich in gas any AM exchange will be preferentially between the (dissipative and therefore cold) gas and the stars, rather than the kinematically hot dark matter, because the amount of AM exchange which occurs depends on the velocity dispersion of the material (as well as the bar strength, and the density of the material; Athanassoula 2003). The effective forces produced by the bar instability act to drive gas inwards from co-rotation (the point at which stars in the disk rotate with the same speed as the pattern speed of the bar) to the central regions. This gas looses its angular momentum which is transfered to the stars in the bar. Interestingly, the forces outside co-rotation may also act to inhibit inflow of gas from the outer regions of the disc, so that gas inflow of external gas onto a disc galaxy is inhibited in the presence of a strong bar (Combes 2008).
 
 One possible conclusion of these theoretical considerations is that strong bars may not be long lived in the presence of significant quantities of gas in a disc galaxy. Numerical simulations generally support this picture of fragile bars and/or bars being unable to grow in the presence of significant amounts of disc gas (e.g. Friedli \& Benz 1993; Berentzen et al. 2007; Heller et al. 2007, Villa-Vargas et al. 2010), although the timescales and gas fractions required are still debated. Some studies suggest that it is the growth of the central mass concentration due to the inflow of gas which is causing bars to dissolve (Debattista et al. 2006, Berentzen et al. 2007; Villa-Vargas et al. 2010), but central mass concentrations must be very large to weaken a bar (e.g. Shen \& Sellwood 2004, Athanassoula et al. 2005); it has also been suggested the gas inflow alone (along with the corresponding increase of AM in the stars in the bar) causes a bar to self-destruct (Athanassoula 2003; Combes 2008).
   
  In this paper, we look for correlations between the (atomic hydrogen) gas content and the likelihood of disc galaxies containing a bar using a sample of 2090 local disc galaxies with both bar classifications and measurements of gas content. Such a test has never been done before on this scale - similar studies have focussed on the details of gas inflow on single galaxies, or small samples of galaxies (e.g. Davous \& Contini 2004; Giordani et al. 2012). Davoust \& Contini (2004) made HI observations of a sample of 144 barred and 110 unbarred Seyfert and star-bursting galaxies finding that the barred galaxies in their sample had lower HI mass fractions than the unbarred galaxies.
  We improve on this sample size by over an order of magnitude, and extend it to include all types of disc galaxies. This allows us to study not just the properties of all barred galaxies together but consider trends with other galaxy properties like stellar mass and colour, which also give clues to the longer term impact of bars on the evolution of disc galaxies. 
  
  To construct the sample we use morphological classifications of bars made by citizen scientists as part of the Galaxy Zoo project (Lintott et al. 2008, 2011)\footnote{www.galaxyzoo.org}, which have previously been used to study the dependence of bar fraction on galaxy properties (Masters et al. 2011) and environment (Skibba et al. 2012), and were also used as the basis of a sample in which bar lengths were measured by citizen scientists and correlations between bar length and other galaxy properties were measured (Hoyle et al. 2011). We combine this morphological data with information on the neutral hydrogen (HI) content of a complete sample of galaxies in the high Galactic latitude Arecibo sky observed as part of the Arecibo Legacy Fast ALFA (Arecibo L-band Feed Array) survey (ALFALFA; Giovanelli et al. 2005). Specifically we use the 40\% of ALFALFA which was recently released as $\alpha40$ by Haynes et al. (2011).  
  
 Where required, we assume a standard cosmological model with $\Omega_m = 0.3$, $\Omega_\Lambda= 0.7$ and $H_0 = 70$ \kmsMpc.
 
\section{Sample and Data}

The Sloan Digital Sky Survey (SDSS) has imaged over one quarter of the sky using a dedicated 2.5m telescope (Gunn et al. 2006) and mosaic CCD Camera (Gunn et al. 1998). Its Main Galaxy Sample (MGS) is a highly complete $r$-band selected sample of galaxies in its Legacy Imaging area which were targeted for spectroscopic followup \citep{MGS}. 

 \begin{figure}
\includegraphics[width=84mm]{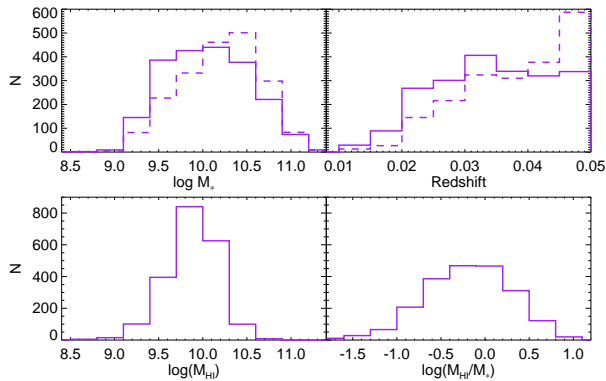}
\caption{Distribution of GZ2 galaxies detected (solid line) or undetected (dashed lines) by ALFALFA. The histograms show stellar mass, redshift, HI mass and HI gas fraction (HI mass per stellar mass). The top right panel (redshift) illustrates that at the higher redshift end of our volume limit more galaxies are undetected in HI - this is a sample bias present in our sample. The tendency for more massive (in $M_\star$) galaxies to be more likely to be undetected is a real feature of the galaxy population. \label{sample}}
\end{figure}

We use a volume limited sub-sample of the MGS galaxies which were included in the second phase of the Galaxy Zoo project (GZ2). We limit the sample to $z<0.05$ in order (1) to have sufficient angular resolution to detect large scale bars over the whole sample, and (2) to remove the frequency range where the San Juan Airport Radar limits ALFALFA's sensitivity to redshifted HI. In addition we use a lower limit of $z>0.01$ to reduce the impact of peculiar velocities on distance errors. The GZ2 sample containing approximately the brightest quarter of the MGS, was selected on $m_r<17.0$, so the volume limit to $z=0.05$ is $M_r<-19.73$ (or $M_r-5\log h < -18.96$). 

We need to identify bars from GZ2 classifications, so limit the sample to those galaxies where at least a quarter of classifiers saw a disc which was not completely edge-on (and therefore answered a question about the presence of a bar). We apply a cut on the axial ratio to remove inclined disks for which bar identifications will be unreliably determined. This cut is $\log (a/b)> 0.3$ (where $\log (a/b)$ is from the exponential $a/b$ measured in the SDSS $r$-band). This effectively limits objects to moderately inclined or face-on disk galaxies with $i<60^\circ$. This cut will make galaxies in our sample easier than average to detect in ALFALFA by reducing the observed HI width (and correspondingly increasing the peak flux of the line). The total Galaxy Zoo bar sample size is 12956 galaxies.
  
  We use a cross match with the ALFALFA 40\% data (Haynes et al. 2011; hereafter $\alpha40$) which provides HI data for 15044 galaxies to $cz=18000$km/s ($z=0.06$) in four patches of the Arecibo sky. Limiting this to a redshift of $z=0.05$ to remove the range of redshifts where RFI from the San Juan Airport radar is at the frequency of redshifted HI removes 1562 HI sources. 
 Only the central portion of the $\alpha40$ area (\ie ~RA=7.5h--16.5h in the North Galactic Cap) overlaps with the SDSS Legacy area. Limiting both HI and optical samples to this region of the sky results in a cross match between 9633 HI sources in $\alpha40$ and 4089 Galaxy Zoo identified fairly face-on disc galaxies ($\alpha40$ covers about 25\% of the SDSS DR7 Legacy area). We use the cross match between $\alpha40$ and the SDSS MGS presented in Haynes et al. (2011) to find 2090 galaxies in common between these two samples - a HI detection rate in our volume limited GZ2 bar sample of 51\%. 
 
\begin{figure}
\includegraphics[width=84mm]{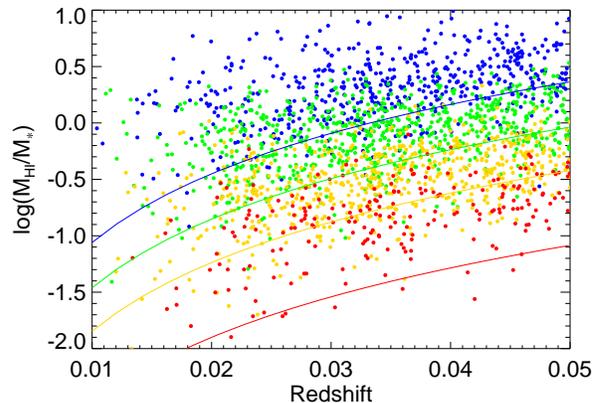}
\caption{The relationship between the range of HI mass fraction and redshift in our sample of 2090 galaxies in a volume limit of $0.01<z<0.05$. The sample consists of galaxies with stellar masses $9.0<\log (M_\star/M_\odot) < 11.5$, and points are colour coded by their stellar mass: {\it blue}:   $\log (M_\star/M_\odot) < 9.7$;  {\it green}:  $9.7<\log (M_\star/M_\odot) < 10.2$; {\it orange}:  $10.2<\log (M_\star/M_\odot) < 10.7$; {\it red}: $ 10.7< \log (M_\star/M_\odot) < 11.5$. The curves show the average HI mass fraction sensitivity of ALFALFA as a function of redshift for the upper mass limit of each subsample (see Section 2.3).  This plot illustrates the extent of HI mass fraction incompleteness as a function of redshift in the sample. \label{HIcomplete}}
\end{figure}

 A limitation of our study is that ALFALFA is optimized for low redshift, low HI mass galaxies, while we study optically bright disc galaxies with morphologies from GZ2. In particular, at the higher redshift end of the volume limit ($z=0.05$), only the most massive galaxies in HI will be detectable\footnote{Figure 3 of $\alpha40$ shows the limiting HI mass as a function of distance. For reference it is $\log (M_{HI}/M_\odot) = 9.6$ at $z=0.05$, and $\log (M_{HI}/M_\odot) = 8.0$ at $z=0.01$}. Figure \ref{sample} show the distribution of detections (solid lines) and non-detections (dashed lines) in our sample as a function of stellar mass, redshift, HI mass and gas fraction (HI mass per stellar mass). This illustrates that at higher redshift end of the sample, more galaxies are undetected in HI. This figure also illustrates that galaxies with higher stellar masses are more likely to be HI poor (high fraction of non-detections among more massive $M_\star$ galaxies). Our sample is volume limited in $r$-band, and is approximately complete for galaxies at all stellar masses between $9.0<\log (M_\star/M_\odot) < 11.5$ at all redshifts, but the HI mass fraction completeness decreases with redshift: at $z=0.05$ we only detect massive galaxies ($\log (M_\star/M_\odot) > 10.2$) with gas fractions $\fHI<1.0$, and no galaxies with $\fHI<0.1$ (see Figure \ref{HIcomplete}). 
   
 \subsection{Photometric Data from SDSS}
 
  \begin{figure*}
\includegraphics[width=160mm]{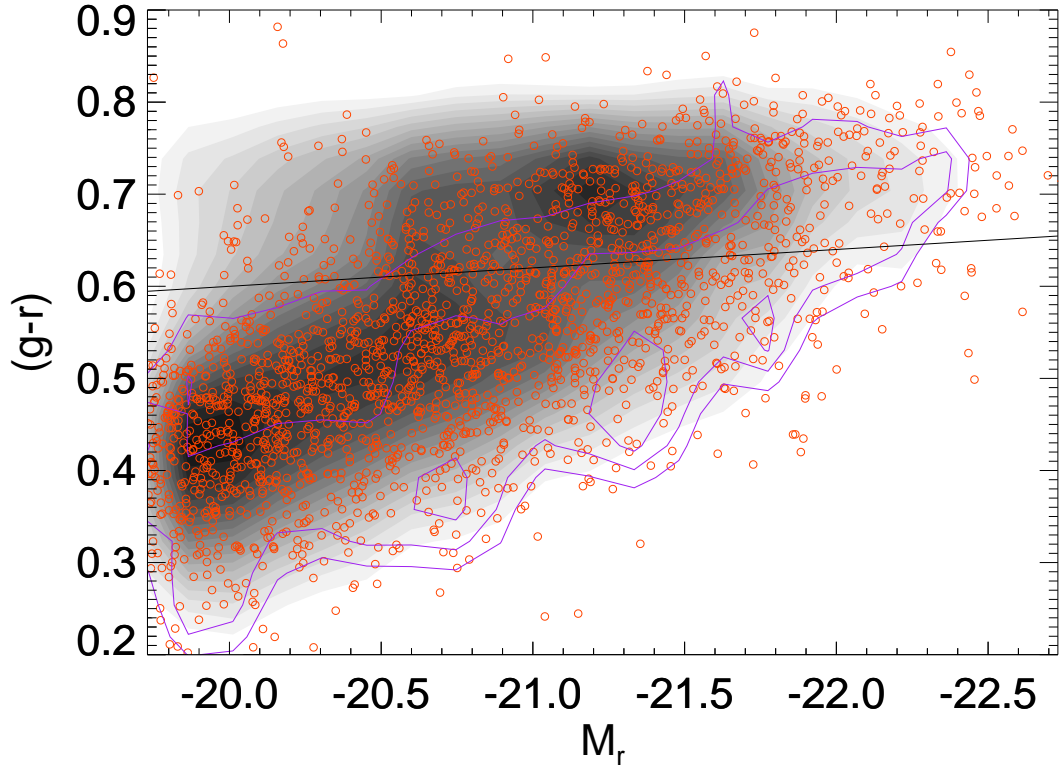}
\caption{A colour-magnitude diagram of all moderately face-on disc galaxies in the GZ2 sample which are in the part of the sky observed by $\alpha40$ (greyscale contours showing the classic blue cloud, red sequence bi-modality). Superimposed are the colours and magnitudes of the HI detected galaxies. The line contours indicate 40\%, 60\% and 80\% detection rates showing that blue sequence spirals are preferentially detected.  We also show the line used by Masters et al. (2010b) to define the blue edge of the ``red sequence", namely $(g-r) = 0.67 - 0.02(M_r + 22)$, based on a fit to Galaxy Zoo identified early-types, which demonstrates that while HI detection in red sequence spirals is rare, it does occur.  \label{CM}}
\end{figure*}

 All photometric data is taken from the SDSS Data Release 7 \citep[DR7][]{DR7}. The reader is referred to York et al. (2000) and Stoughton et al. (2002) for details on the hardware, software and data-reduction in SDSS. Photometry was taken in 5 bands: $ugriz$ (Fukugita et al. 1996). For total magnitudes, we use the Petrosian magnitudes (Petrosian 1976, Strauss et al. 2002), while colours are calculated from the model magnitudes (with the aperture set in the $r$-band). In addition, we use axial ratio information (from the $r$-band exponential model fit), and spectroscopic redshifts. All photometric quantities are corrected using the standard Galactic extinction corrections (Schlegel, Finkbeiner \& Davis 1998), and a small k-correction (to $z=0$) is applied (Blanton et al. 2003; Blanton \& Roweis 2007). 
 
 Stellar masses are estimated using the colour dependent stellar mass-light ratio based on SDSS $i$-band magnitudes and $(g-i)$ colours presented in \citet{Z09}. Specifically this means we use $\log M_\star/L_i = -0.963 + 1.032(g-i)$ from Zibetti et al. (2009) together with a solar magnitude of $M_{\odot,i} = 4.58$ (Blanton et al. 2001).  For our sample of nearly face-on normal disc galaxies this technique should result in a reasonable estimate of the stellar mass, with typical uncertainty of 0.2-0.3 dex, dominated by the uncertainty in the estimate of the stellar mass-to-light ratio (Zibetti et al. 2009). 
 
  We show in Figure \ref{CM} an optical $(g-r)$  colour magnitude diagram of all galaxies in the GZ2 bar sample which are in the part of the sky observed by ALFALFA (greyscale contours). Superimposed are the colours and magnitudes of the HI detected galaxies in the sample, with the line contours indicating a 40\%, 60\% or 80\% detection rate. We show the line used by Masters et al. (2010b) to define the blue edge of the ``red sequence" (to identify ``red spirals"). As expected, HI detection rates increase towards the bluer and lower luminosity part of the diagram, however we note with interest that a significant number of Galaxy Zoo 2 identified disc galaxies which are in the ``red sequence" are still detected in HI. We remind the reader that our sample excludes very inclined disk galaxies which could be reddened by dust (e.g. Masters et al. 2010a), so these are intrinsically ``red spirals". This has previously been observed in $\alpha40$ considering the full cross match with SDSS (e.g. Haynes et al. 2011), and also seen by Toribio et al. (2011a,b).  
   
 \subsection{Morphologies from Galaxy Zoo 2}
 
Morphological classifications from GZ2 are based on information provided by multiple independent citizen scientists. The median number of citizen scientists classifying each galaxy in GZ2 is 45. Before reaching the question about  bars, each volunteer must answer two questions. These are {\it ``Does the galaxy have features or a disc?"}, and {\it ``Is the galaxy totally edge-on?"} (see Figure 1 of Masters et al. 2011). We include in our sample only those galaxies for which the weighted $p_{\rm features}p_{\rm not edge-on} > 0.25$ thus requiring that the number of answers to the bar question is at least 25\% of the total number of people classifying the galaxy. In addition, in order to reduce the impact of erroneous classifications we require that at least ten people answered the question {\it ``Is a bar visible in the galaxy?"}; the median number is $N_{\rm bar} = 30$. We call the weighted fraction\footnote{We follow a similar weighting procedure as described in Lintott et al. (2008) to reduce the impact of extremely divergent classifications.}  of these classifiers answering that they see a bar, to the total number of classifiers answering the question, the bar ``probability", or $\pbar = N_{\rm bar, yes}/N_{\rm bar}$. 

 In Appendix A we compare the GZ2 bar classifications to other independent classifications of bars (both from visual inspection and ellipse fitting). This comparison confirms that visual inspection by multiple citizen scientists can reliably identify bars in galaxies. It demonstrates that GZ2 bar identified by $\pbar>0.5$ are similar to the classic strong bar (SB) classification. In addition our comparison suggests that galaxies with $0.2<\pbar<0.5$ can be identified as weakly barred galaxies, while truly unbarred galaxies will have $\pbar<0.2$. 
 
 In the rest of the paper we sometimes refer to GZ2 strong bars simply as ``bars", and galaxies with $\pbar<0.5$ which may host a weak bar as ``unbarred". Detailed comparisons between our work and other samples should recall the precise definition given here. 

 \begin{figure*}
\includegraphics{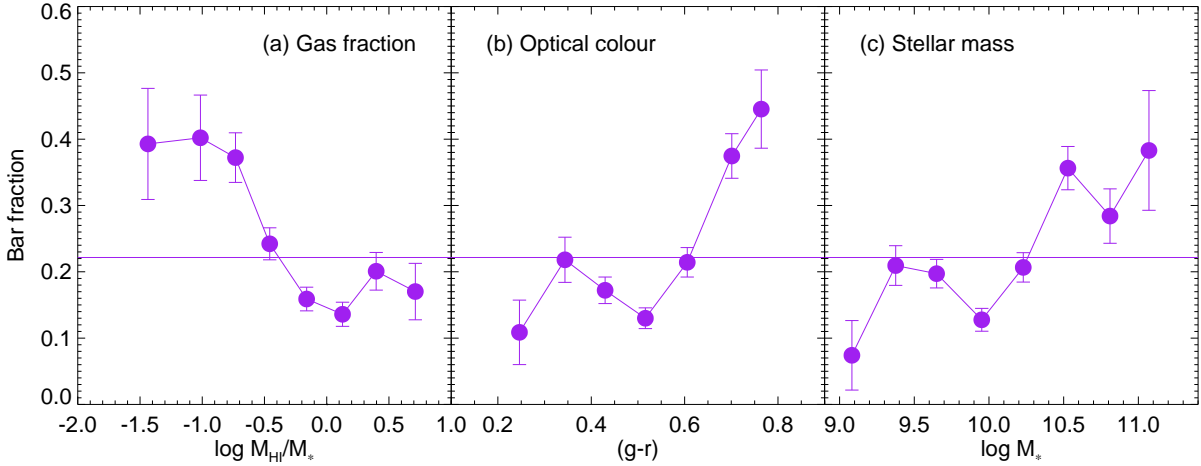}
\caption{The strong bar fraction as a function of (a) gas fraction, (b) optical $(g-r)$ colour and (c) stellar mass for 2090 Galaxy Zoo galaxies detected in HI by $\alpha40$. Strong bars are identified from GZ2 classifications using $\pbar > 0.5$ (as discussed in Section 2.2). Figure 4 shows that the strong bar fraction increases as atomic gas content decreases and as optical colour and stellar mass increase. The errors shown are Poisson counting errors on the fractions -- these are underestimates for the fractions close to zero (\ie ~very gas rich, and blue galaxies, see Cameron 2011). The horizontal lines show the strong bar fraction for all HI detected galaxies of $22\pm1$\%. Galaxies undetected in HI in the sample have a strong bar fraction of $32\pm1$\%.  \label{barfraction}}
\end{figure*}

  \subsection{HI Data from ALFALFA}
  
  The ALFALFA survey was initiated in 2005 following the commissioning of a 7 pixel feed array working at 21cm (or L-band) on the Arecibo Telescope (ALFA, or Arecibo L-band Feed Array). Full details of the plans for ALFALFA can be found in Giovanelli et al. (2005). Its goal is to survey 7000 deg$^2$  of high Galactic latitude sky observable with the Arecibo Telescope. ALFALFA is sensitive to HI lines in the redshift range of  $v=-1600$--18000 \kms ($z<0.06$).   
 As discussed in Haynes et al. (2011) the detectability of a HI source by ALFALFA depends both on the integrated HI line flux, and the width of the HI profile. The limiting sensitivity for the catalogued HI sources \footnote{In this study we are always considering HI matched to optical counterparts, therefore make use of both Code 1 and Code 2 sources from $\alpha40$.}  is measured from the data by Haynes et al. (2011) as,
  \begin{eqnarray}
  \log S_{\rm lim} = & 0.5 \log W_{50} - 1.23  & \log W_{50}< 2.5, \nonumber \\
                                    & \log W_{50} - 2.49  & \log W_{50}\geq 2.5,
 \end{eqnarray}
  (where $S_{\rm lim}$ is the limiting flux in Jy \kms, and $W_{50}$ is the width of the HI line in \kms). This limiting sensitivity refers to the average properties of $\alpha40$ and cannot be used to give a detection limit for a specific galaxy, but is useful to give an idea of the regions of parameter space in $M_{\rm HI}$ versus other galaxy properties which are undetectable by ALFALFA. 
  
HI masses are calculated from the total HI flux observed by ALFALFA using the standard conversion (e.g. Haynes et al. 2011) of  
 \be
 M_{\rm HI} = 2.356\times 10^5 D_{\rm Mpc}^2 S_{\rm{Jy km/s}}.
 \ee
The typical 1$\sigma$ error on this (including distance errors) is estimated to be 0.15-0.25dex for galaxies in the mass range of our sample (Figure 19 of Haynes et al. 2011). 

We use the usual definition of the HI gas fraction relative to the stellar mass of the galaxy $f_{HI} = M_{\rm HI}/M_{\star}$. Adding the errors on $M_{\rm HI}$ and $M_{\star}$ to obtain a typical uncertaintly on $\log \fHI$ of 0.25-0.4 dex gives an upper limit to the error of this distance independent quantity, as both $M_{\rm HI}$ and $M_{\star}$ error estimates contain the error on the assumed distance. In what follows we sometimes refer to $\fHI$ simply as the gas fraction, even though it is the atomic hydrogen gas fraction. We comment on the impact on our results of possible hidden molecular H$_2$ at the end of Section 4. 

 For a galaxy with stellar mass $M_\star$ the minimum gas fraction which can be observed is 
  \be
  f_{\rm HI,lim}= 2.356\times 10^5 D_{\rm Mpc}^2 S_{\rm lim} /M_\star.
 \ee
 We use this to estimate the limiting HI gas fraction as a function of stellar mass at the redshift limits of our sample. This estimate includes an assumption about the typical width of observed HI emission at a given stellar mass\footnote{Based on the detected galaxies we use $W_{50,\rm{max,obs}} = 320 + 220 (\log (M_\star/M_\odot) - 10)$ \kms.}.  At the lower redshift limit, this gives a detectable HI fraction which ranges from $\fHI = 0.002$--0.3  for $\log (M_\star/M_\odot) = 11.5$ and $9.0$ respectively, while at the upper redshift limit the minimum detectable HI fraction is $\fHI = 0.06$--7.0. 
    
\section{Results}

\subsection{Bar Fraction with Gas Fraction}
 In Figure \ref{barfraction} we observe a clear anti-correlation between the strong bar fraction and HI mass fraction of disc galaxies in our sample. This suggest that either (1) HI rich galaxies are less likely to host strong bars, or (2) strongly barred galaxies have lower atomic gas fractions than unbarred/weakly barred galaxies. The median gas fraction among barred galaxies in our sample is $\fHI=0.39$ (with an inter-quartile range, or IQR of 0.19--1.1, or expressed in $\log$ space, $\log \fHI = -0.40^{+0.43}_{-0.33}$), compared to a median value of $\fHI=0.74$ (IQR 0.35--1.5, or $\log \fHI = -0.13^{+0.30}_{-0.32}$) in unbarred galaxies.   

While HI poor galaxies are preferentially detected in the near part of our sample (Section 2.3), the observed trend cannot be explained by resolution effects.  The SDSS images used to identify bars have a median physical resolution of 1.3 kpc at $z=0.05$ which is sufficient to detect all galactic scale bars across the whole redshift range. 

We also confirm with this sample (middle and right panel of Figure \ref{barfraction}) the previously observed trends of higher bar fraction in disc galaxies with higher stellar masses and redder optical colours (e.g. Nair \& Abraham 2010b; Masters et al. 2011; Skibba et al. 2012). We note that as a set of galaxies selected to have been detected in HI this sample is biased towards lower mass, ``blue cloud" late-type (small bulge) spirals than previous Galaxy Zoo studies of the bar fraction (Masters et al. 2011; Skibba et al. 2012), which also use a more luminous volume limit to $z=0.06$.

 While the trends for more strong bars to be found in massive, optically red and gas poor disc galaxies is the most obvious feature of the plots in Figure 4, it can also be seen that a small peak in strong bar fraction is seen in lower mass ($\log (M_\star/M_\odot) < 10.0$), bluer, and more gas rich galaxies. That the trends of bar fraction are not monotonic across the Hubble sequence and seems to have a minimum at around $\log (M_\star/M_\odot) = 10.0$ has been noted previously (e.g. in the RC3: Odewahn 1996, and more recently, Nair \& Abraham 2010b, Masters et al. 2011), and most likely indicates a difference in evolution for bars in different mass galaxies. 

\begin{figure*}
\includegraphics[height=220mm,angle=-90]{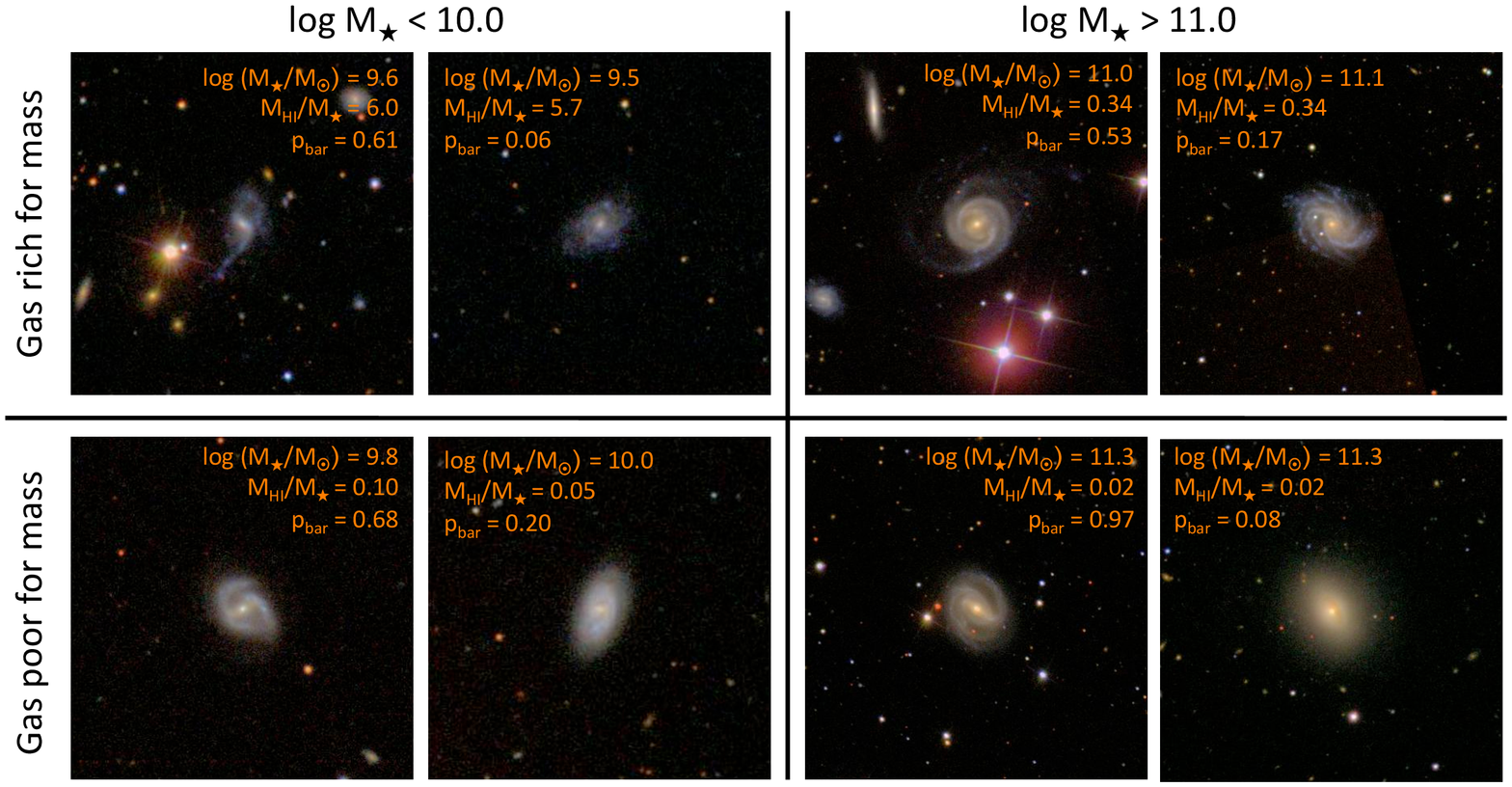}
\caption{Example $gri$ colour cutouts of galaxies in the sample. The galaxies are chosen to show extremes in low/high stellar mass and gas rich/gas poor galaxies for both barred and unbarred objects and are circled in both panels of Figure \ref{masshistograms}. The stellar mass, gas fraction, and the fraction of GZ2 classifiers identifying a bar ($\pbar$) are indicated on each panel. The galaxies are: top row: (1) gas rich low mass barred galaxy:  SDSS J152957.41+072650.5; (2) gas rich low mass unbarred galaxy: SDSS J122244.60+135755.4; (3) gas rich high mass barred galaxy:  SDSS J141057.23+252950.0; (4) gas rich high mass unbarred galaxy: SDSS J155323.17+115733.0; bottom row: (1) gas poor low mass barred galaxy:  SDSS J122350.72+040513.7; (2) gas poor low mass unbarred galaxy: SDSS J122630.20+080339.3 ; (3) gas poor high mass barred galaxy:  SDSS J161403.28+141655.6; (4) gas poor high mass unbarred galaxy: SDSS J125455.16+272445.7. Images are shown scaled to the Petrosian radii (the width in arcseconds is 10$r_p$) as they would have been seen by citizen scientists on the GZ2 website. 
  \label{examples}}
\end{figure*}

Example images, of high and low stellar and HI mass galaxies with and without bars are shown in Figure \ref{examples} \footnote{More example images can be see at {\tt http://www.icg.port.ac.uk/$\sim$mastersk/GZ\_ALFALFAImages}.}. 

\subsection{Breaking Degeneracies with Gas Content, Stellar Mass and Colour}

\begin{figure*}
\includegraphics[width=160mm]{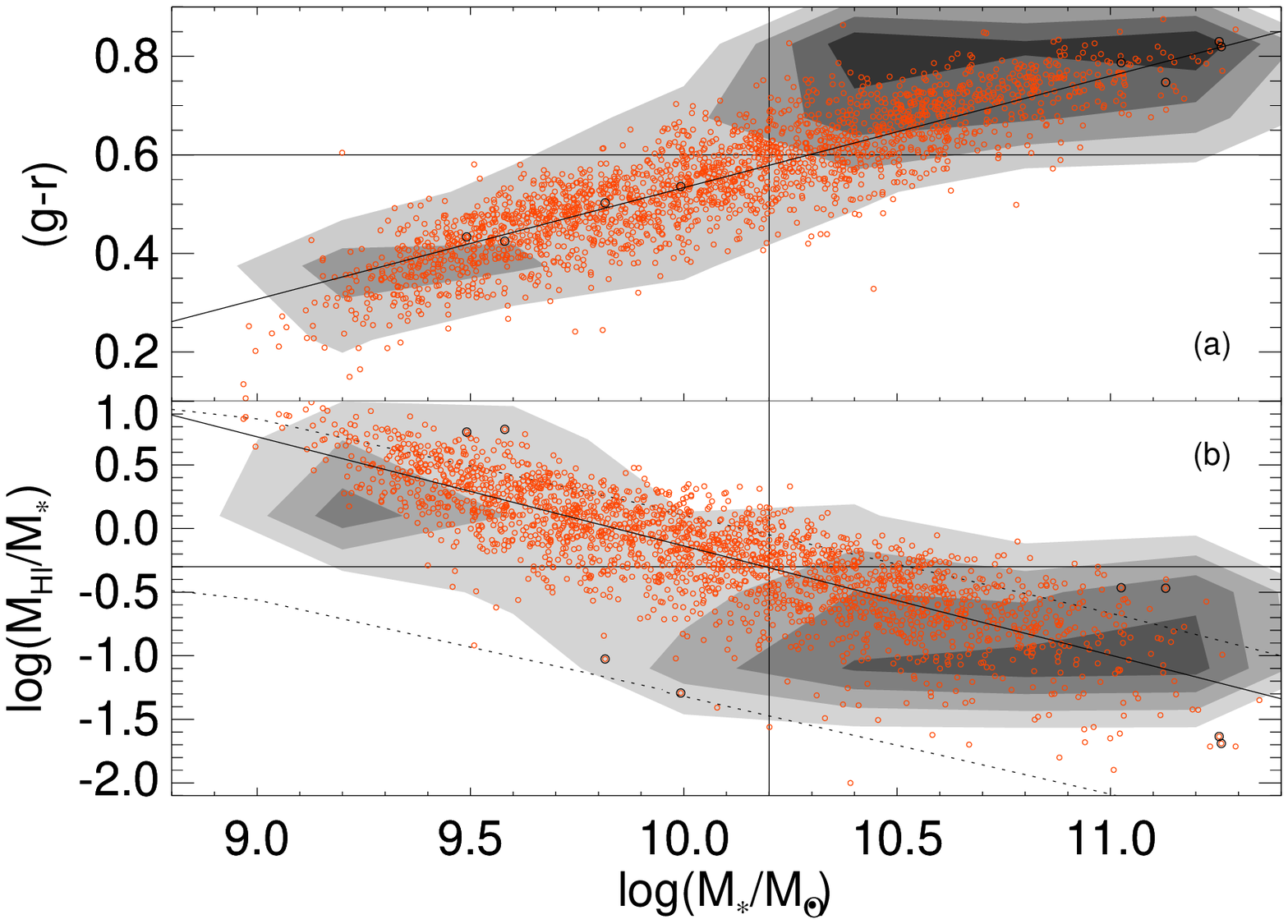}
\caption{The relationship between (a) stellar mass and optical colour; and (b) stellar mass and HI mass fraction for the 2090 galaxies in our sample. Points in both panels show the location of galaxies as a function of gas fraction, colour and stellar mass. Fits to the observed trends are shown (diagonal solid lines). The contours hi-light the strong bar fraction on the plot (ie. the fraction of galaxies with $\pbar>0.5$ from GZ2, see Section 2.2), with greyscale indicating 10\%, 20\%, 30\% and 40\% strong bars. This plot demonstrates that more massive disc galaxies are redder and have lower gas fractions (points), and also that the bar fraction is highest among massive disc galaxies which are redder and have lower gas fractions than is typical for their stellar mass (contours). We indicate on the plot cuts used later to make subsets of the sample. The vertical line shows a mass cut of $\log (M_\star/M_\odot) = 10.2$, the typical transition mass between disc and bulge dominated populations, while horizontal lines indicate: (a) a colour of $(g-r)=0.6$ which roughly splits blue and red disc galaxies; (b) $\log(M_\star/M_{HI}) = -0.30$ (or $M\star = 2 M_{\rm HI}$) which we will use to split ``gas rich" and ``gas poor" galaxies in Section 3.3 below. The dotted lines in panel (b) show an estimate of the limiting gas fraction which would make galaxies at the lower and upper redshift range of our sample detectable by ALFALFA (as discussed in Section 2.3). Finally the circled points indicate galaxies whose images are shown as examples of bar classifications in Figure \ref{examples}. \label{masshistograms}}
\end{figure*}

It is well known (e.g. most recently seen in ALFALFA data by Toribio et al. 2011a,b, Catinella et al. 2010, Fabello et al. 2011, Huang et al. 2012) that the atomic gas content of galaxies correlates with both stellar mass and optical colour, which are of course also correlated via the colour-magnitude relation. We illustrate these correlations in Figure \ref{masshistograms} showing the locations of HI detected galaxies in our sample as a function of stellar mass, gas fraction and $(g-r)$ colour. The best fit to the trends are shown as solid lines. 

Given these correlations and the fact that the strong bar fraction increases towards higher stellar mass, redder disc galaxies (Nair \& Abraham 2010b, Masters et al. 2011, Skibba et al. 2012) we must ask if all, or part, of the correlation between gas fraction and bar fraction can be explained by the combination of the correlations between gas fraction and stellar mass and those between stellar mass/colour and bar fraction. 
 
 The bar fraction is indicated in Figure \ref{masshistograms} by the grey scale contours which show strong bar fractions of between 10-40\%. From this we observed that the bar fraction peaks most strongly among the higher stellar mass disc galaxies which are both redder and less gas rich than is typical for their stellar mass. This already demonstrates that the correlations between gas fraction and stellar mass/colour cannot explain the full increase of bar fraction with decreasing gas fraction.    
 
\subsubsection{Bar Fraction with Gas Deficiency}

In this section we will use the relationship between stellar mass and gas fraction observed in Figure \ref{masshistograms} to calculate the expected gas fraction for a galaxy of a given stellar mass. We find a trend of 
\be \langle \log (M_{\rm HI}/M_\star) \rangle = -0.31 - 0.86(\log (M_\star/M_\odot) - 10.2) \ee 
with a typical scatter of $\sigma_{\log (M_{\rm HI}/M_\star)} = 0.27$ dex. Clearly the selection function plays a role in shaping the trends, and will reduce the observed scatter by preferentially removing gas poor galaxies at a given stellar mass. However, we point out that the deeper HI observations of the GASS survey (GALEX-Arecibo SDSS Survey) which targeted galaxies with $M_\star > 10^{10} M_\odot$; (Catinella et al. 2010) demonstrate that there are few galaxies at $10.0<\log (M_\star/M_\odot) < 10.5$ with gas fractions below 10\%, and the observed trends are similar to those we see here. \citet{T11} have also previously studied the typical HI content of isolated disc galaxies in $\alpha40$ and also find similar correlations to us. Finally, \citet{F11} use a stacking technique to place limits on the HI content of undetected early-type galaxies (selected using optical concentration) and find similar trends of the HI gas fraction with stellar mass. 

Knowing the expected gas fraction for a given stellar mass is important, as we can use it to define a measure of HI deficiency, e.g. as was used in Haynes \& Giovanelli (1984), Solanes et al. (2001), Toribo et al. (2011a,b), Cortese et al. (2011) and study trends of the bar fraction with this quantity. We define our version of HI deficiency as 
\be \rm{HI}_{\rm def,\star} = \langle \log (M_{\rm HI}/M_\star) \rangle -  \log (M_{\rm HI}/M_\star).\ee

\begin{figure}
\includegraphics[width=84mm]{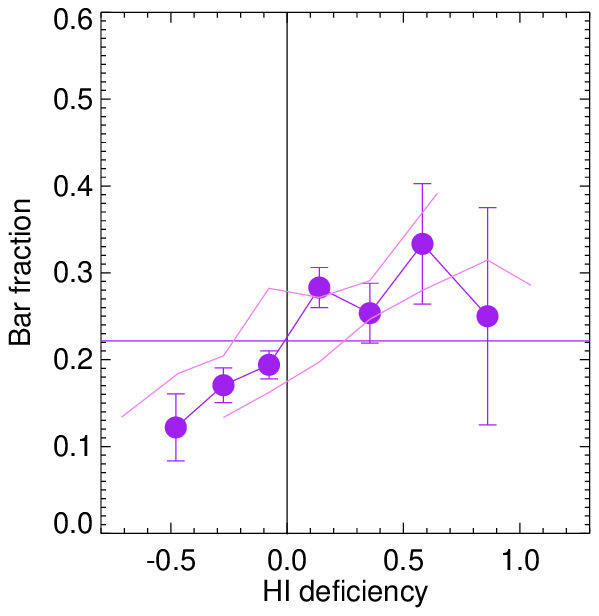}
\caption{The strong bar fraction as a function of gas deficiency for 2090 Galaxy Zoo galaxies detected in HI by $\alpha40$. Gas deficiency is calculated relative to the typical gas fraction for galaxies with the same stellar mass (HI$_{\rm def,\star}$ in Eqn 5), so a gas deficiency of zero (vertical line) is the typical gas fraction for a given stellar mass (see Figure 6). This figure demonstrates that galaxies at a given stellar mass are more likely to be found hosting a bar if they are gas deficient. The errors on the points show Poisson counting errors.  The paler lines show the same result for HI deficiencies calculated as the trend in Figure \ref{masshistograms} now $\pm$0.27 dex (the observed scatter in HI deficiency). As expected, this moves the trend of bar fraction with gas deficiency to the right or left by approximately 0.27 dex.  The y-range of this plot is identical to Figure \ref{barfraction} for ease of comparison. \label{barfraction_gasdef}}
\end{figure}

 The trend of bar fraction with HI deficiency for GZ2 disc galaxies detected by $\alpha40$ is shown in Figure \ref{barfraction_gasdef}. This Figure demonstrates that galaxies which have more HI gas than is usual for their stellar mass are less likely to be observed with a strong bar than average:  down to a bar fraction of 13$\pm4$\% in the most gas rich. Those galaxies which have less HI gas than is usual for their stellar mass are more likely to be observed with a strong bar: up to a bar fraction of 36$\pm11$\% in the most gas deficient (combining the last two bins in Figure 7). 
 
Equivalently, barred galaxies in our sample are found to be HI deficient; with a median HI$_{\rm def,\star}=0.05\pm0.01$ dex; and unbarred galaxies are HI rich for their mass with a median HI$_{\rm def,\star}=-0.04\pm0.01$ dex.  We estimate the significance of these differences using our measured 1$\sigma$ range of HI deficiency of HI$_{\rm def,\star}=0.27{\rm dex}/\sqrt N$. 
 
 \subsubsection{Splitting the sample by Stellar Mass and Colour}
 
\begin{figure}
\includegraphics[width=84mm]{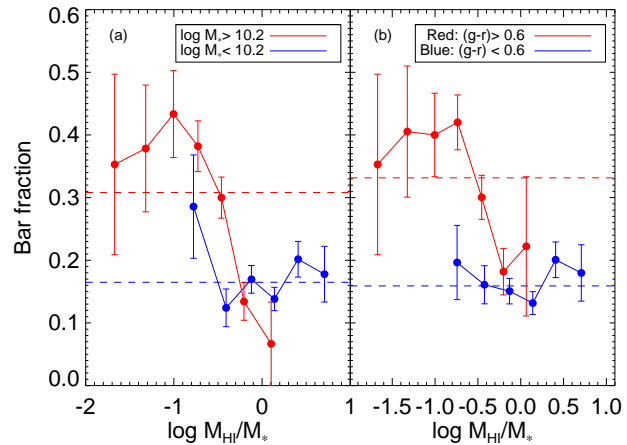}
\caption{The strong bar fraction as a function of gas fraction for 2090 GZ2 galaxies detected in HI by $\alpha40$ split into 2 subgroups by either (a) stellar mass or (b) optical colour. As in Figure \ref{barfraction}, the horizontal lines show the overall bar fraction for each sub-sample. We see that at a fixed gas fraction a redder disc galaxy is more likely to be found hosting a bar (or barred discs are redder),  however at a fixed gas fraction the stellar mass of the disc galaxy correlates less well with whether or not it hosts a bar.  \label{split1}}
\end{figure}

 Another way to address possible degeneracies between gas fraction, stellar mass, galaxy colour and the probability of hosting a bar is to split the sample by these properties and look for residual trends. Figure \ref{split1} shows the trend of (strong) bar fraction with gas fraction split now into: (a) high and low stellar mass disc galaxies or (b) red and blue disc galaxies. We use $\log (M_\star/M_\odot) = 10.2$ as the dividing line between the high and low mass subsamples, as suggested by many studies who give a similar mass division to where the properties of galaxies seem to change (e.g. Strateva et al. 2001; Kauffmann et al. 2003; Baldry et al. 2004). To split the sample into red and blue disc galaxies we use $(g-r)=0.6$ as the divider between the red sequence and blue cloud.  A summary of bar fractions in different subsets of the data is given in Table 1. 
 
\begin{table}
\caption{Summary of (strong) bar fractions for Subsets of the Data \label{barfractions}}
\label{class}
\begin{tabular}{lcc}
\hline
Sample & N & Strong bar fraction \\
\hline
HI non-detections & 1999 &  $32\pm1\%$ \\
\hline
All HI detections & 2090 & $22\pm1\%$ \\
~~~~~High mass ($\log (M_\star/M_\odot) > 10.2$) & ~821 & $31\pm2$\%\\
~~~~~Optically red ($(g-r)>0.6$) & ~757 & $33\pm2$\%\\
~~~~~Gas poor ($\fHI < 2.0$) & ~836 & $31\pm2$\% \\
~~~~~Gas deficient  (HI$_{{\rm def},\star} > 0.0$) & ~953 & $27\pm2$\% \\
~~~~~Low mass ($\log (M_\star/M_\odot) < 10.2$) & 1268 & $16\pm1$\%\\
~~~~~Optically blue ($(g-r)<0.6$)  & 1333 & $16\pm1$\%\\
~~~~~Gas rich ($\fHI > 2.0$) & 1254  & $16\pm1$\% \\
~~~~~Gas rich for mass (HI$_{{\rm def},\star} < 0.0$) & 1137 &$18\pm1$\% \\
\hline
\end{tabular}
\end{table}
 
 In the sample split by stellar mass, we observe the increase of (strong) bar fraction with stellar mass between the two subsets, from $16\pm1$\% for the low mass subsample to $31\pm2$\% for the high mass. However at a fixed gas fraction there is no statistical difference between the high and low stellar mass subsets, except at around $\fHI=0.3$ where high mass galaxies have a bar fraction of $30\pm3\%$, while low mass have $12\pm3\%$. 
 
Within the high mass subset there remains a strong correlation of bar fraction with gas fraction; once galaxies have at least a 10\% HI gas fraction the probability that they will host a strong bar is observed to drop to equal, or even below that seen in low mass disc galaxies (down to $7\pm7$\% for the most gas rich of the high mass subset \ie ~15 galaxies with $M_{HI} \sim 1.3 M_\star$, only 1 of which is barred). 

Interestingly, among the lower mass galaxies there remains only a mild residual trend between bar fraction and gas fraction, only significant in the lowest gas fraction bin in the subset, which has a bar fraction of $29\pm8\%$ (at $\fHI\sim0.1$). However we note that almost all low mass galaxies which could be detected by $\alpha40$ have $f_{\rm gas}\geq0.1$, so this sample does not constrain the behaviour of low mass disc galaxies containing little gas. We notice (Figure \ref{masshistograms}) that for $\log (M_\star/M_\odot) < 10.2$, the sample contains only five galaxies with $f_{\rm gas}<0.1$, two of which host strong bars, (see two of them in lower left of Figure \ref{examples}). This is consistent with an increase in bar fraction for such galaxies (ie. to 40\%, but with Poisson error of $\pm$28\%), however this clearly would need to be tested using a larger sample of low stellar mass HI poor galaxies.

 The split between the bar fraction in the red and blue disc galaxy sub-samples (left panel of Figure \ref{split1}) is slightly larger: the bar fraction is $33\pm2$\% for red discs versus $16\pm1$\% for blue discs. Red disc galaxies are observed to be more likely to host strong bars than blue disc galaxies at all gas fractions. A residual trend with gas fraction is still observed within red disc galaxies. Once a red disc galaxy has at least 10\% HI (relative to its stellar mass), the probability it will host a strong bar is observed to fall off, down to $19\pm4$\% in the most gas rich red disc galaxies with $f_{HI}\sim1$ and which have bar fractions consistent with the blue disc galaxy subsample. 
 
 However in the blue disc galaxy subsample we observe no significant residual trend of bar fraction with gas content, but note that the majority of the blue discs in our sample have $f_{\rm gas}>0.1$. 
  
In summary, even for red, massive disks we see a residual correlation between (strong) bar fraction and atomic gas content. Once red, massive disc galaxies have $f_{\rm gas}>0.1$ they are increasingly unlikely to host a bar. However among blue, low mass galaxies (which in our sample are dominated by relatively gas rich galaxies with $f_{\rm gas}>0.1$) strong bars are unlikely at all (observed) gas fractions.

 \subsubsection{Splitting the sample by Atomic Gas Content}

\begin{figure}
\includegraphics[width=84mm]{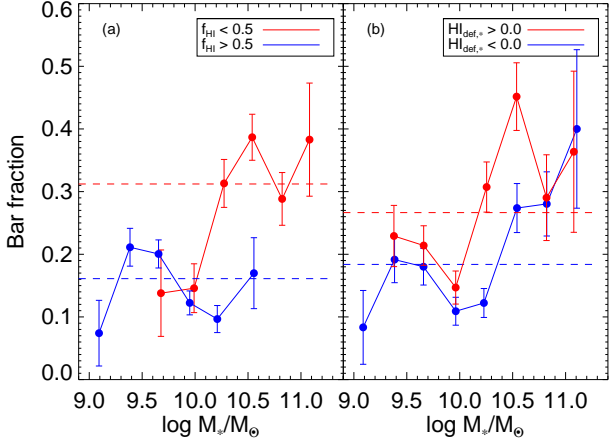}
\caption{The strong bar fraction as a function of stellar mass  for 2090 Galaxy Zoo galaxies detected in HI by $\alpha40$ split into (a) gas rich ($f_{\rm HI}>0.5$) or gas poor ($f_{\rm HI}<0.5$ galaxies; and (b) gas deficient (HI$_{{\rm def},\star}>0.0$) or gas rich for mass (HI$_{{\rm def},\star}<0.0$). The horizontal lines show the bar fraction for the whole of each sub-sample.  Note that gas poor galaxies fall out of our sample at the lowest stellar masses of our sample, while gas rich massive galaxies are intrinsically very rare.  We see that much of the correlation between bar fraction and stellar mass disappears when the sample is split into gas rich and gas poor. The biggest difference in bar fraction with gas content occurs in the intermediate mass range ($10<\log (M_\star/M_\odot) < 10.7$), while very (stellar) massive, or low mass galaxies in our sample display no correlation between gas content and bar fraction. \label{barfraction_mstar_gasfraction}}
\end{figure}

In Figure  \ref{barfraction_mstar_gasfraction} we show the correlations of strong bar fraction with stellar mass in subgroups split by the gas fraction (at $\log(\fHI) = -0.30$, or $M\star = 2 M_{\rm HI}$, as suggested by the typical gas fraction of a green valley and transition mass galaxy (\ie ~$(g-r)\sim0.6$ and $\log (M_\star/M_\odot) = 10.2$; see Figure \ref{masshistograms}), or by HI gas deficiency as defined above. 

Averaged over the full stellar mass range of our sample, the (strong) bar fraction in the gas poor subsample is $31\pm2$\% significantly higher than the (strong) bar fraction for the gas rich subsample ($16\pm1$\%), while the gas deficient subsample has a bar fraction of $27\pm2$\%, compared to $18\pm1$\% in those galaxies with more gas than is typical for their stellar mass. 

In the left hand panel we observe that the residual correlation of bar fraction with stellar mass is rather flat across the whole range of stellar masses probed by the gas rich disc galaxies, and for $\log (M_\star/M_\odot) > 10.2$ in the gas poor subsample. Lower mass galaxies ($\log (M_\star/M_\odot) < 10.2$) seem to have the same low bar fraction independent of gas content, while the bar fraction in more massive discs depends strongly on gas content. 

The right panel of Figure \ref{barfraction_mstar_gasfraction} displays the residual correlation of bar fraction with gas content calibrated to the typical gas content at a given stellar mass (i.e. gas deficiency as defined in Eqn. 5, and see the lower panel of Figure \ref{masshistograms}). In both gas deficient and gas rich galaxies there remains a residual correlation of bar fraction to increase with stellar mass. The two trends are statistically indistinguishable for $\log (M_\star/M_\odot) < 10.0$ or $\log (M_\star/M_\odot) >10.7$; the main correlation between HI gas deficiency and bar fraction happens in the intermediate mass ranges which represent the transition region between blue cloud and red sequence discs ($10<\log (M_\star/M_\odot) < 10.7$). This is also apparent in the lower panel of Figure \ref{masshistograms} as a ``banana shape" of the contours hi-lighting the bar fraction as a function of both gas mass and stellar mass.  Here we split the sample at HI$_{{\rm def},\star}=0.0$. We also try using HI$_{{\rm def},\star}=0.3$ as the dividing line (as suggested by the 1$\sigma$ scatter in the trend observed in Figure 6) and find no qualitative difference in the results. 

In summary splitting the sample by gas content, either as an absolute HI fraction, or using the HI deficiency parameter defined in Section 3.2.1 demonstrates that the main difference in bar fraction with gas content happens in the intermediate mass disc galaxies which populate the transition region (sometimes called the ``green valley") of the colour magnitude diagram. If a galaxy in that region is gas rich it is much less likely to host a bar (or if barred it is less likely to be gas rich).  Lower mass galaxies tend to have low bar fraction regardless of gas content, and higher mass galaxies tend to have high bar fractions regardless of gas content.

\section{Discussion}

The main result of this paper is that the strongly barred disc galaxies from Galaxy Zoo 2 are observed to be more likely to be HI gas poor than unbarred or weakly barred disc galaxies (Figure \ref{barfraction}), or equivalently HI poor galaxies are more likely to be observed hosting a bar than HI rich galaxies. This correlation cannot be explained by redder, more massive disc galaxies being simultaneously more likely to host a strong bar and more likely to be HI poor. We demonstrate that at a fixed stellar mass there remains significant residual correlations of bar fraction with gas content, particularly over the stellar mass range $10<\log (M_\star/M_\odot) < 10.7$ which corresponds to the typical mass of a disc galaxy in the ``green valley". 

In this section we consider the causal links between atomic gas content and bar fraction in disc galaxies that may create the observed correlations. We consider three possible explanations: (1) Bars in disc galaxies cause atomic gas to be used up more quickly; (2) Increasing the atomic gas content in a disc galaxy either causes bars to form more slowly, or to self-destruct more quickly; (3) Bar fraction and gas content are both driven by correlation with environmental effects. 

\begin{enumerate}
\item {\it Bars in disc galaxies cause atomic gas to be used up more quickly}:  \\ There is clear physical explanations, and observational evidence, that bars in gas rich disc galaxies funnel gas into the central regions of the galaxy where it is turned into molecular gas and eventually forms stars (e.g. Ho et al. 1997, Sheth et al. 2005, Ellison et al. 2011, Lee et al. 2012). This mechanism will accelerate the globally averaged atomic gas consumption, by concentrating the gas, and in addition, by removing gas from the outer regions of the disc, should cause those regions to cease forming new stars and become optically red (in the absence of external gas inflow). The timescale for the transfer of gas along a bar has been estimated at $\sim10^8$ years (Athanassoula, priv. comm to Coelho \& Gadotti 2011). The only previous study of the HI gas fraction of barred galaxies (Davoust \& Contini 2004) concluded that the decrease in HI content they observed in 113 barred starburst and Seyfert galaxies was due to this mechanism. Here we have confirmed this result in a larger and more representative sample of normal disc galaxies. 
 
\item {\it Gas in disc galaxies inhibits bar formation}: \\ Bars are dynamical systems. Many theoretical studies (both analytical and using numerical simulations) have shown that over time bars will evolve by the exchange of angular momentum between the bar, disc and halo material (e.g. Ostriker \& Peebles 1973, Sellwood 1980, Athanassoula 2002, Berentzen et al. 2006 and see the review in Sellwood 2010). Due to its dissipative nature, a cold gas component can exchange angular momentum with stars in the bar very efficiently (Athanassoula 2003), and therefore is predicted to have important effects on the evolution of a bar. 

Numerical simulations of galaxies with gas, generally find that the presence of a gaseous component will (1) inhibit the formation of bars, by dampening the initial bar instability (Berentzen et al. 1998, Villa-Vargas et al. 2010), and (2) eventually destroy the bar, either by building up a central mass concentration (CMC) which destroys the bar orbits (Friedli \& Benz 1993, Shen \& Sellwood 2004, Athanassoula et al. 2005, Debattista et al. 2006, Berentzen et al. 2007), or by the process of transferring angular momentum from the gas to the bar (Bournaud \& Combes 2002, Bournaud et al. 2005). The timescales for dissolution, and the amount of gas required to significantly affect the bar evolution depend on the details of the simulation. Timescales vary from 1-2.5 Gyr (e.g. Bournard \& Combes 2002, who saw multiple periods of bar formation and dissolution in a 20 Gyr simulation, Bournaud et al. 2005, Heller et al. 2007) to much longer (5 Gyr or more in Athanassoula et al. 2005, Berentzen et al. 2006), while the gas fraction required varies from a few percent of the visible matter (e.g. Freidli \& Benz 1993, Bournaud et al. 2005) to as much as 20\% (Shen \& Sellwood 2004, Debattista et al. 2006). 

In many simulations however, the gas fractions investigated are lower that those observed in real disc galaxies. This may be partly due to an inability to stimulate bar growth in model galaxies with large gas fractions (as discussed in Villa-Vargas et al. 2010).  In Berentzen et al. (2007) it is claimed that most disc galaxies have typically less than 10\% of their disc mass in stars (no citation given), however recent studies of the HI atomic gas content of galaxies indicate the fraction is usually much larger, and at the stellar mass investigated by Berentzen et al. (2007) the {\bf average} gas fraction ($M_{HI}/M_\star$) of disc galaxies is observed to be about 8\% with a range between the detection limit of about 3\% and as much as 60\% \citep{C10}. Both Debattista et a. (2006), and Villa-Vargas et al. (2010) ran simulations with up to 50\% cold gas (by mass in the disc), although as we show in Figure \ref{barfraction} (and see e.g. Toribo et al. 2010, Catinella et al. (2010) observationally disc galaxies are seen to have even just atomic gas masses up to 10 times larger than their stellar mass (ie. having 90\% of the baryonic mass of the disc in atomic gas). To fully understand correlation between bar formation and gas content will require a greater range of disc gas fractions to be simulated.

Fortunately, due to ongoing improvement in computing power, and increasing interest in the impact of secular evolution of galaxies we can expect more and more detailed simulations of bars in disc galaxies containing gas, including live halos embedded in full cosmological simulations. 

\item {\it The correlation between bar fraction and gas content is driven by their mutual dependence on environment}: \\ Disc galaxies do not live in isolation. It has long been suggested that at least some bars may be triggered by environmental interactions (e.g. Byrd et al. 1986, Noguchi 1996, Berentzen et al. 2004). It has recently been shown that even after correcting for stellar mass and colour, there is some residual tendency for barred disc galaxies to cluster more strongly than unbarred disc galaxies, particularly on scales of $\sim 400$ kpc (Skibba et al. 2012), a subtle effect previously unobserved in smaller samples (e.g. Li et al. 2009, Aguerri et al. 2009, M{\'e}ndez-Abreu, S{\'a}nchez-Janssen, \& Aguerri 2010, Mart{\'{\i}}nez \& Muriel 2011). In addition it is well known that the atomic gas content of galaxies is reduced as they enter virialized structures (e.g. Haynes \& Giovanelli 1984, Solanes et al. 2001, Toribio et al. 2011b). While ram pressure stripping is probably the dominant process removing gas (see e.g. the review of Boselli \& Gavazzi 2006), even quite gentle processes such as strangulation/starvation can remove halo gas in lower density environments (Larson, Tinsley \& Caldwell 1980, Balogh, Navarro \& Morris 2000, Bekki et al. 2002).  Galaxy harassment in clusters may also act, both forming a stellar bar, and driving the gaseous component to the centre of the galaxy (Moore et al. 1996,1998). All together this suggests that disc galaxies suffering even quite mild environmental effects may both be more likely to have triggered bar instabilities, and lower than average amounts of atomic gas. Finally, there is some evidence that bars are more likely to be triggered by environmental effects when the disc is depleted of gas (Berentzen et al. 2004).  More simulations of the internal structure of disc galaxies which include both stars and gas, and are embedded in full cosmological simulations (e.g. Heller et al. 2007) will help to explain these issues better. 

\end{enumerate}

If bars are transitory objects on the timescale of less than 1-2 Gyr or so (e.g. Block et al. 2002, Combes 2008, van den Bergh 2011) our observations suggest that gas content is driving the likelihood of a disc galaxy being observed with or without a bar. That at a fixed gas fraction there is no (little) correlation between bars and stellar mass is easily explained in this case, since over the lifetime of the galaxy during which it has been building up its stellar mass the galaxy could have multiple periods of hosting strong, weak, or no bar, and the overall correlation between stellar mass and bar fraction is driven by more massive disc galaxies having lower gas fractions. The observation that at a fixed gas fraction/stellar mass bars are more likely to be found in redder discs can also be explained. Neglecting the effects of dust, discs will be optically red (in $(g-r)$ filters) if they have had no significant star formation in the last 0.5 Gyr or so (e.g. the model SEDs in Figure 4 of Schawinski et al. 2007) a time scale comparable to the lifetime of a strong bar in the models of Bournaud \& Combes (2002) and Bournaud et al. (2005). Environmental correlations with bars (e.g. Skibba et al. 2012) are also easy to explain - for example, if strangulation like processes remove the outer gas from disc galaxies, while the inner gas is funneled to the centre along the bar this would quickly clear the disc of gas causing any bar present in the galaxy to become extremely long lived (e.g. Athanassoula 2003), and star formation to effectively cease in all but the very central regions. 

If bars are very long lived, then disc galaxies without strong bars have simply not yet developed them. The gas content could still be driving observed correlations of the bar fraction with galaxy properties through its ability to diminish the bar instability (Berentzen et al. 1998) and inhibit bar growth, but perhaps more likely the secular evolution driven by the bar will cause atomic gas to be used up more quickly in barred galaxies.

We see hints that if a gas rich galaxy does (unusually) host a strong bar, it is likely to be optically redder than a similar gas rich galaxy without a bar. 
We propose that this effect could be due to the exchange of angular momentum beyond co-rotation acting to inhibit infall of external gas (e.g. Combes 2008). While this effect may not be strong enough to entirely shut down the inflow of atomic gas from the halo of a disc galaxy, we suggest that it might act to slow it down, and in this way help to regulate the star formation in such galaxy (in a process which in some ways is similar to strangulation type mechanisms). Consistent with this idea, we observe that barred disc galaxies are typically redder in the region interior to bar co-rotation than exterior to it (Hoyle et al. 2011), and that the four luminous and strongly barred spirals with resolved HI imaging in THINGS (The HI Nearby Galaxy Survey; Walter et al. 2008) all have significant HI holes in the bar region (but we note that many unbarred galaxies do too). This idea is also in agreement with Wang et al. (2012), who use a sample of massive ($M_\star > 10^{10} M_\odot$) SDSS galaxies with bars identified from ellipse fitting methods, along with a comparison of central and global specific star formation rates, to conclude that bars may play a role in quenching global star formation in massive disc galaxies, at the same time as increasing central star formation (see also e.g. Ellison et al. 2011, Coehlo \& Gadotti et al. 2011, Oh et al. 2012, Lee et al. 2012). This picture of ``bar quenching" could and should be tested by more detailed numerical simulations of bars in gas rich galaxies, and by resolved HI imaging of a larger sample of gas rich strongly barred galaxies. Curiously it has the same sort of mass scaling used in semi-analytic models of galaxy formation to introduce the star formation feedback more usually tied to the presence of AGN (which like strong bars are more common in more massive disc galaxies). 

We finish by reminding the reader that atomic HI is not the only kind of gas which is found in galaxies. Also likely important for bar dynamics is the molecular H$_2$. This gas phase might actually be expected to be more important than HI, since the H$_2$ in disc galaxies is more likely to be found inside the disc radius to which the bar reaches, while a significant fraction of the HI in discs is usually found outside the optical radius (as first systematically shown by Broeils and Rhee 1997), and HI distributions often show central holes (e.g. Roberts 1975, Bosma 1978, Shostak 1978). H$_2$ is also important as the phase which plays the direct role in star formation and therefore gas consumption. HI must condense to H$_2$ in molecular clouds before stars can form. 

The HI to H$_2$ ratio can vary substantially from galaxy to galaxy, by as much as two orders of magnitude in spiral galaxies (from $M_{{\rm H}_2}/M_{\rm HI} = 0.03-3.0$; Boselli et al. 2002, Lisenfeld et al. 2011, Saintonge et al. 2011a). These observational studies appear to find that the median value decreases slightly from massive early type spirals to later types. It was observed to be as much as $\langle M_{{\rm H}_2}/M_{\rm HI} \rangle = 0.8$ in S0s (but with a high uncertaintly) by Lisenfeld et al. (2011), while Saintonge et al. (2011a) observe $\langle M_{{\rm H}_2}/M_{\rm HI} \rangle = 0.3 $ in massive discs ($\log (M_\star/M_\odot) >10.0$), and $\langle M_{{\rm H}_2}/M_{\rm HI} \rangle= 0.1-0.15$ is typical for later type spirals (Boselli et al. 2002, Lisenfeld et al. 2011). 

Despite the importance of the molecular gas phase in understanding galaxy evolution, observational samples are relatively small due to the difficulties of obtaining the CO measurements used to estimate H$_2$ content. None of the samples is large enough to study correlations of bar fraction with molecular gas content in the way we have done here, although a search for systematic variation in the $M_{{\rm H}_2}/M_{\rm HI}$ between barred and unbarred discs couple be done with the available data. Saintonge et al. (2012) see some suggestion of a reduced molecular gas depletion time for barred galaxies, but that sample is very small. We also note that the molecular gas content of a galaxy can be estimated via its correlation with other more commonly measured quantities (e.g as discussed in Boselli et al. 2002, Biegel et al. 2008, Saintonge et al. 2011b). An interesting extension of the observations presented in this paper would be to use something like this to add estimates of molecular gas content to the correlations we observe with atomic hydrogen. 

\section{Summary}

We use a sample of optically detected SDSS MGS galaxies with morphological classifications from the Galaxy Zoo project, and data on their HI content from the ALFALFA survey to consider correlations between the bar fraction and atomic gas content of disc galaxies. Our sample includes all moderately face-on disc galaxies with bar classifications from GZ2 in a volume limit of $0.01<z < 0.05$ ($M_r<-19.73$ for GZ2 selection), which were in addition detected in HI by the ALFALFA survey (specifically the $\alpha40$ release, Haynes et al. 2011). The stellar mass range is $9.0<\log (M_\star/M_\odot) < 11.5$, and the HI mass limit ranges from $\log M_{\rm HI}/M_\odot = 8.0$ at $z=0.01$, to $\log M_{\rm HI}/M_\odot = 9.6$ at $z=0.05$. We define a strongly barred galaxy as one which has a GZ2 bar probability of $\pbar>0.5$, galaxies with $\pbar<0.5$ are considered unbarred, or weakly barred. We find: 
\begin{itemize}
\item There is a significant correlation between observed strong bar fraction and gas content such that bar fraction increases as gas content decreases. Barred disc galaxies contain less atomic hydrogen on average than unbarred disc galaxies. In addition we see the correlation between bar fraction and stellar mass or optical colour previously observed in a similar sample by Nair \& Abraham (2010b), Masters et al. (2011). 
\item Using a HI gas deficiency parameter (how much more or less HI gas a galaxy has relative to the typical value for its stellar mass) we show that there is a significant correlation between HI deficiency and bar fraction such that HI deficient galaxies are more likely to host a bar, or barred galaxies are more likely to be HI deficient.
\item Using subsets of the sample split into massive/low mass; red/blue and gas rich/gas poor (see Figures \ref{split1} and  \ref{barfraction_mstar_gasfraction}) we observe that at a fixed stellar mass gas poor galaxies have more bars than gas rich ones. At a fixed gas fraction it is (optically) redder disc galaxies that are most likely to host bars, with less dependence on stellar mass. The biggest difference in bar fraction with gas content appears at the mass scale of typical ``green valley" galaxies. 
\item Finally we see hints that if a gas rich galaxy does (unusually) host a strong bar, it is likely to be optically redder than a similar gas rich galaxy without a bar. 
\end{itemize}

We discuss three possible causal relationships which can explain these observations: (1) that bars in disc galaxies cause atomic gas to be used up more quickly, (2) that increasing the atomic gas content in a disc galaxy either causes bars to form more slowly, or to self-destruct more quickly, and (3) that bar fraction and gas content are both driven by correlation with environmental effects. Depending on the galaxy in question, all of these mechanisms may work together to create the observed correlations. Further study, including information on environment on a sample of disc galaxies with atomic gas content and bar identifications will be needed to draw stronger conclusions.  

As numerical simulations of the growth of structure in the Universe become more and more complex and probe a wider dynamic range of physical scales, understanding how the internal properties of galaxies affect their global star formation histories will become more and more important. The Galaxy Zoo project provide invaluable, reliable and reproducible information on the morphologies of galaxies in samples large enough to study the complicated intercorrelations which drive galaxy evolution.

\paragraph*{ACKNOWLEDGEMENTS.} 

This publication has been made possible by the participation of more than 200,000 volunteers in the Galaxy Zoo project. Their contributions are individually acknowledged at \texttt{http://www.galaxyzoo.org/volunteers}. Galaxy Zoo 2 was developed with the help of a grant from The Leverhulme Trust.  KLM acknowledges funding from The Leverhulme Trust as a 2010 Early Career Fellow. RCN acknowledges STFC Rolling Grant ST/I001204/1 "Survey Cosmology and Astrophysics". CJL acknowledges support from an STFC Science in Society fellowship. Support for the work of K.S. was provided by NASA through Einstein Postdoctoral Fellowship grant number PF9-00069, issued by the Chandra X-ray Observatory Center, which is operated by the Smithsonian Astrophysical Observatory for and on behalf of NASA under contract NAS8-03060.

We thank the many members of the ALFALFA team who have contributed to the acquisition and processing of the ALFALFA data set over the last six years. The ALFALFA team at Cornell is supported by NSF grants AST-0607007 and AST-1107390 to RG and MPH and by a grant from the Brinson Foundation.

We acknowledge helpful discussions with Michael Williams, O. Ivy Wong, and the S$^4$G collaboration bar working group, particularly Kartik Sheth and Bruce Elmegreen. We also thank Fabio Barazza for making available an electronic file of bar identifications from ellipse fitting method as described in Barazza et al. (2008). The anonymous referee significantly aided the publication process by providing an extraordinarily fast, and helpful referee report.

The Arecibo Observatory is operated by SRI International under a cooperative agreement with the National Science Foundation (AST-1100968), and in alliance with Ana G. MŽndez-Universidad Metropolitana, and the Universities Space Research Association.

Funding for the SDSS and SDSS-II has been provided by the Alfred P. Sloan Foundation, the Participating Institutions, the National Science Foundation, the U.S. Department of Energy, the National Aeronautics and Space Administration, the Japanese Monbukagakusho, the Max Planck Society, and the Higher Education Funding Council for England. The SDSS Web Site is http://www.sdss.org/. 

The SDSS is managed by the Astrophysical Research Consor- 
tium for the Participating Institutions. The Participating Institu- 
tions are the American Museum of Natural History, Astrophysical 
Institute Potsdam, University of Basel, University of Cambridge, 
Case Western Reserve University, University of Chicago, Drexel 
University, Fermilab, the Institute for Advanced Study, the Japan 
Participation Group, Johns Hopkins University, the Joint Institute 
for Nuclear Astrophysics, the Kavli Institute for Particle Astro- 
physics and Cosmology, the Korean Scientist Group, the Chinese 
Academy of Sciences (LAMOST), Los Alamos National Labora- 
tory, the Max-Planck-Institute for Astronomy (MPIA), the Max- 
Planck-Institute for Astrophysics (MPA), New Mexico State Uni- 
versity, Ohio State University, University of Pittsburgh, University 
of Portsmouth, Princeton University, the United States Naval Ob- 
servatory and the University of Washington.

\appendix
\section{Bar identification from Galaxy Zoo 2 Compared to Other Methods}

We have compared bar identifications in a GZ2 bar sample similar to the one described in Section 2 (and used in Masters et al. 2011; Skibba et al. 2012; namely 15292 galaxies in a volume limit of $0.017<z<0.06$, $M_r < -20.15$, which are not more edge on that $i\sim 60^\circ$ and which have reliable bar identifications from GZ2) with other published bar identifications. The largest cross match comprising 3638 galaxies comes from the sample of visual classifications performed on SDSS $gri$ images by Preethi Nair (Nair \& Abraham 2010a; hereafter NA10). In addition we find classic classifications for 557 galaxies in our sample from the Third Reference Catalogue of Bright Galaxies (RC3; de Vaucouleurs et al. 1991), and 243 galaxies also have bar classifications using the method of ellipse fitting on SDSS images (as described in Barazza et al. (2008; hereafter B08) and kindly shared for the purposes of comparison by Fabio Barazza).  

\begin{figure}
\includegraphics[width=9cm]{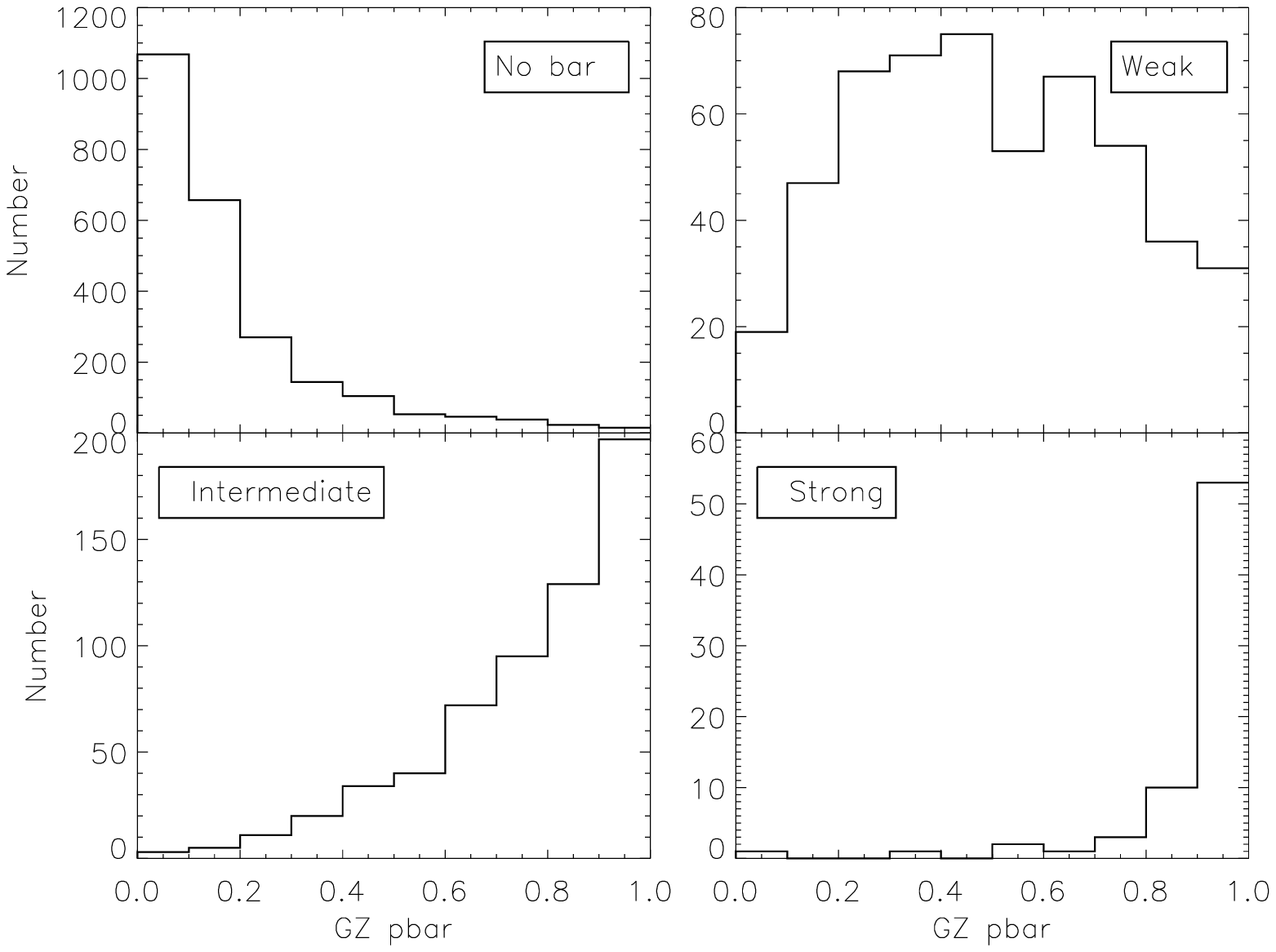}
\caption{The distribution of GZ2 bar likelihoods for the four subsamples of bar classifications given by NA10. Top left: 2418 galaxies identified as not having a bar by NA10; top right: 521 galaxies identified as having a weak bar by NA10; bottom left: 606 galaxies identified as having an intermediate bar by NA10; bottom right: 71 galaxies identified as having a strong bar by NA10. \label{Nair}}
\end{figure}

Table \ref{barID} shows the comparison between these different bar classifications and thresholds of $\pbar$ discussed in this paper. In addition we show in Figure \ref{Nair}, histograms of the distribution of $\pbar$ from GZ2 for the 4 classes of bar ID provided by NA10. Both the Table and Figure demonstrate that the agreement between GZ2 and NA10 bar IDs is very good. Strong and intermediate bar from NA10 almost all ($>90\%$) have the high values of $\pbar>0.5$ used in this paper (and previous GZ2 work) to identify strong bars. In addition galaxies without bars from NA10 all have low values of $\pbar$ (71\% with $\pbar<0.2$; or 92\% with $\pbar<0.5$) while weak bars from NA10 have intermediate values of $\pbar$ (prompting the description in this paper of galaxies with $0.2<\pbar<0.5$ as weak barred systems). 

The agreement between GZ2 and RC3 bar IDs is also good, particularly in identifying galaxies without bars. Most RC3 weak/unsure bars have very low values of $\pbar$. There are a also a small but not insignificant number of RC3 strong barred galaxies with low values of $\pbar$. We find that these are typically also classified as unbarred by NA10 and therefore suggest that the mismatch is either due to (1) the images used in RC3 being more sensitive to subtle bar features than the SDSS images used by GZ2 and NA10; or (2) human error in the RC3. 

The comparison between GZ2 and B08 bar IDs again demonstrates that the methods agree well at identifying galaxies without bars. However, a significant fraction of B08 identified bars have low values of $\pbar$ (66\% with $\pbar<0.5$ and even 37\% with $\pbar<0.2$). We attribute this difference to B08 bar IDs including both weak and strong bars, while the GZ2 selects only strong bars (Barazza et al. priv. comm, based on the size distribution of bars identified by B08 and GZ2).  

\begin{table*}
\caption{Comparison between GZ2 bar IDs and other classifications \label{barID}}
\label{class}
\begin{tabular}{lcccccc}
\hline
\hline
& NA10 no bar & NA10 no bar  & RC3 no bar & RC3 weak/unsure bar & B08 no bar \\
& & $M<10^{10}M\sun$\\
& $N=2418$ & $N=463$ & $N=309$ & $N=69$ & $N=113$ \\
\hline
$\pbar <0.2$ & 71\% (1725) & 75\% (346)  & 60\% (184) & 45\% (31)  & 80\% (90) \\
$\pbar<0.5$ & 92\% (2225) & 96\% (446) & 83\% (256) & 68\% (47) & 96\% (109) \\
\hline
\hline
& NA10 weak bar & NA10 weak bar  \\
& & $M<10^{10}M\sun$\\
\hline
& $N=521$ & $N=90$ \\
$0.2<\pbar<0.5$ & 40\% (206) &51\% (46) & \\
$0.2<\pbar<0.8$ &  74\% (388) &78\% (70)  & \\
\hline
\hline
& NA10 int bar & NA10 int bar  & NA10 strong bar & NA10 strong bar & RC3 strong bar & B08 bar \\
& & $M<10^{10}M\sun$ & & $M<10^{10}M\sun$\\
\hline
& $N=606$& $N=71$ & $N=58$& $N=2$ & $N=179$ &$N=130$ \\
$\pbar > 0.2$ & 99\% (598) & 99\% (70)& 97\% (56) & 100\% (2) & 76\% (136) & 63\% (82) \\
$\pbar > 0.5$ & 90\% (545) & 97\% (69)& 76\% (44) & 100\% (2)  & 58\% (104) & 34\% (44) \\
\hline
\hline
\end{tabular}
\end{table*}

This comparison between GZ2 and other published bar classifications demonstrates the power of citizen science methods for visual classification. As the GZ2 $\pbar$ value is the result of lots of pairs of independent (and fresh) eyes it does not make spurious mistakes. Expert classification clearly helps with the details and tricky cases, but those tricky cases must involve subjective decisions, and in addition can be prone to human error (something as simple as hitting the wrong key). Automatic classifications are quantitative, but can be prone to being influence by the unexpected. Where ten or more citizen scientists independently classify a galaxy and most see a bar, we can be very certain that something which looks like a bar is present in the image. 
\end{document}